\documentclass[12pt]{article}

\usepackage{latexsym}

\usepackage{graphics}
\usepackage{graphicx}

\begin{document}


\title{Rotating black ring on Kaluza-Klein bubbles}

\author{
    Petya G. Nedkova\thanks{E-mail:pnedkova@phys.uni-sofia.bg}, Stoytcho S. Yazadjiev \thanks{E-mail: yazad@phys.uni-sofia.bg}\\
{\footnotesize  Department of Theoretical Physics,
                Faculty of Physics, Sofia University,}\\
{\footnotesize  5 James Bourchier Boulevard, Sofia~1164, Bulgaria }\\
}

\date{}

\maketitle

\begin{abstract}
We construct a new exact solution to the 5D Einstein equations describing rotating black ring with a single angular momentum surrounded by two Kaluza-Klein bubbles. The solution is generated by 2-soliton B\"{a}cklund transformation. Its physical properties are computed and analyzed. The corresponding static solution, the rotating black string and the boosted black string are reproduced as limits.
\end{abstract}


\sloppy

\section{Introduction}

The stationary axisymmetric solutions to the five dimensional vacuum Einstein equations have been studied intensely in the recent years in both
cases of asymptotically flat and asymptotically Kaluza-Klein spacetimes and a good progress has been achieved. At this stage black solutions
with  a symmetry group $R\times U(1)^2$  are completely classified by the uniqueness theorems \cite{HY1}, \cite{HY2} and general methods for their construction are developed.

The 5D Einstein equations with $R\times U(1)^2$ symmetry group are completely integrable and reduce to a two dimensional non-linear sigma model which allows the application of solitonic techniques.  The most general method for generation of solitonic solutions of the Einstein equations with described symmetries is the inverse scattering method developed by Belisnkii and Zakharov \cite{BZ1}, \cite{BZ2}. A B\"{a}cklund transformation for the two dimensional non-linear problem was also discovered, originally in 4D gravity by Neugebauer \cite{NG}, \cite{SK} and recently adapted to five dimensions \cite{Mishima}. It is relevant only for generation of solutions with a single angular momentum, however, and in this subsector is equivalent to the inverse scattering method \cite{TIM1}. Both solitonic techniques generate solutions  by purely algebraic operations performed on a simpler already known solution, called a seed.

At present many exact solutions of the 5D Einstein equations are available constructed by means of these methods. The Myers-Perry black hole \cite{MyersPerry}, and the black ring, originally obtained by Emparan and Reall \cite{Emparan1}, were subsequently generated using solitonic techniques \cite{Pomeransky}, \cite{IM}, \cite{PSenkov}. More complicated configurations that occur in higher dimensional gravity were constructed as well -  black lens \cite{Teo},  black Saturn \cite{ElvangFigueras}, multi rings \cite{EvslinKrishnan}, \cite{Izumi}, \cite{ElvangRodriguez}, black hole and bubble sequences \cite{IMT}, \cite{TIM2} (see also \cite{Kimura}).

A good progress in constructing exact solutions to the 5D Einstein-Maxwell(-dilaton) gravity and 5D minimal supergravity  was also achieved
and many explicit exact solutions were generated \cite{Y1}-\cite{Y3}, \cite{BCCGSW}-\cite{FJRV}. Black hole uniqueness theorems were also formulated and proven for
certain sectors in  5D Einstein-Maxwell(-dilaton) gravity and 5D minimal supergravity \cite{HY3}-\cite{Y4}.

\paragraph{}The aim of the current paper is to construct a new stationary and axisymmetric solution to the 5D Einstein equations with Kaluza-Klein asymptotics. It describes a rotating black ring on two Kaluza-Klein bubbles and the two-solitonic B\"{a}cklund transformation is applied as a solution generation method. The static solution with the same rod structure was found by Elvang and Horowitz \cite{Elvang}. It was later generalized to sequences of arbitrary (even) number of black holes and bubbles \cite{EHO} and its charged counterparts were generated in \cite{KY}-\cite{YN2}. Related solutions representing two black holes on a Kaluza-Klein bubble, with a single angular momentum along either spacelike Killing field were found in \cite{IMT}, \cite{TIM2}.  In the literature there exist also interesting Kaluza-Klein black hole solutions with an unusual (different from $R^4\times S^1$) asymptotic. For explicit examples we refer the reader to \cite{SSW}-\cite{Y_sq}.

\paragraph{}The article is organized as follows. In section 2 the solution describing static black ring on Kaluza-Klein bubbles is briefly presented. In section 3 the generating technique of its stationary couterpart is discussed,  the seed solution is constructed and the rotating black ring on two Kaluza-Klein bubbles is generated. In section 4 the solution is thoroughly analyzed. The regularity conditions are estimated, physical characteristics such as mass, tension, angular velocity and angular momentum are computed, and Smarr relations are derived. The final section is devoted to some limits of the solution. Such are the static solution of Elvang and Horowitz \cite{Elvang}, the rotating black string \cite{Mishima} and the boosted black string \cite{Emparan2}.

\section{Static solution}

The solution describing a static black ring on Kaluza-Klein bubbles was first derived in \cite{Elvang}, although
it was not interpreted as such. It was obtained following the Emparan and Real higher dimensional generalization \cite{ER} of the Weyl method for constructing static axisymmetric solutions in 4D (also called Weyl solutions). The metric for any D-dimensional spacetime with D-2 commuting orthogonal Killing vectors, one of which timelike, can be written in the following form

\begin{equation}
 ds^2= \sum_i \epsilon_i e^{2U_i} (dx^i)^2 + e^{2\nu}(d\rho^2+dz^2),
\end{equation}
\noindent
where the 2-dimensional surfaces orthogonal to the Killing fields are parameterized in Weyl canonical coordinates, and all the metric functions depend only  on  $\rho$ and $z$.
Emparan and Reall showed that the solution to the Einstein equations in vacuum can be obtained in the following manner. The functions $U_i$, $i= 1,..,N-2$, proved to be solutions to a 3D Laplace equation in flat space

\begin{equation}\label{laplace}
 \frac{\partial^2 U_i}{\partial \rho^2} + \frac{1}{\rho}\frac{\partial
 U_i}{\partial \rho} + \frac{\partial^2 U_i}{\partial z^2} = 0.
\end{equation}

Hence, they are called potentials in resemblance to the Newtonian potentials produced by certain axisymmetric sources. The potentials $U_i$ are not all independent as they must obey a further constraint

\begin{equation}\label{constraint}
 \sum_i U_i = \ln \rho + {\rm constant}.
\end{equation}

Actually, $\ln{\rho}$ is a solution to the Laplace equation corresponding to a Newtonian potential produced by an
infinite rod of zero thickness lying along the $z$-axis. For that reason, the metric functions $U_i$ can be interpreted as Newtonian potentials produced by linear sources along the $z$ axis, which according to the constraint (\ref{constraint}) must add up to an infinite rod. The sources of $U_i$, which can be finite or semi-finite, constitute the so called rod structure. The notion of rod structure can be extended to the case of stationary axisymmetric spacetimes \cite{Harmark}. It is a basic characteristic of stationary axisymmetric solutions, as, according to the uniqueness theorems \cite{HY1}, \cite{HY2}, a vacuum solution is determined completely by its rod structure and angular momenta in the asymptotically flat and asymptotically Kaluza-Klein spacetime\footnote{Strictly speaking the uniqueness theorems of \cite{HY1} and \cite{HY2} are based on the so-called
interval structure. The interval structure is a more precise mathematical definition of the  rod structure  with very important restrictions
which were omitted in the definition of the rod structure. For the aims of the present paper, however,  it is sufficient to use the more physical and simple notion of rod structure rather than the more precise mathematical notion of interval structure.}.

\paragraph{}Once the potential $U_i$ is obtained from (\ref{laplace}), the remaining metric functions $\nu_i(\rho, z)$ are determined by the integrable system

\begin{eqnarray}
\partial_\rho \nu &=& - \frac{1}{2\rho }
+ \frac{\rho}{2} \sum_{i=1}^{D-2} \left[ (\partial_\rho U_i)^2
- (\partial_z U_i)^2 \right]\,,
\\
\partial_z \nu &=& \rho \sum_{i=1}^{D-2} \partial_\rho U_i \partial_z U_i \,.
\end{eqnarray}

Elvang and Horowitz used the method we just described to obtain the following static axisymmetric solution to the 5D vacuum Einstein equations in spacetime with Kaluza-Klein asymptotic

\begin{eqnarray}
g_{tt} &=&  - e^{2U_1} =
               - \frac{R_3 - \zeta_3}{R_2 - \zeta_2}, \\ \nonumber
g_{\phi\phi}   &=&  e^{2U_2} =
                     \frac{(R_2 - \zeta_2)(R_4 - \zeta_4)}
                    {(R_1 - \zeta_1)(R_3 - \zeta_3)}, \\ \nonumber
g_{\psi\psi}   &=& e^{2U_3} = \left(R_1 - \zeta_1 \right)
                        \left(R_4 + \zeta_4 \right), \\ \nonumber
g_{rr}  &=& g_{zz} =
 e^{2\nu} =
               \frac{Y_{14} Y_{23}}{4 R_1 R_2 R_3 R_4}
               \sqrt{\frac{Y_{12} Y_{34}}{Y_{13} Y_{24}}}
               \left(\frac{R_1 - \zeta_1}{R_4 - \zeta_4} \right),
 \end{eqnarray}

In order to write the solution in more convenient form some auxiliary functions are introduced according to

\begin{eqnarray}
  \zeta_1 = z-a ~  ;~~~
  \zeta_2 = z-b ~  ;~~~
  \zeta_3 = z+b ~  ;~~~
  \zeta_4 = z+c \, ,
\end{eqnarray}
\begin{eqnarray}
  &&R_i    = \sqrt{\rho^2 + \zeta_i^2}, \\
  &&Y_{ij} = R_i R_j + \zeta_i \zeta_j + \rho^2,
\end{eqnarray}
where $a$, $\pm b$ and $-c$ are the rod endpoints.

\begin{figure} [h]
\begin{center}
      \includegraphics[width=8.cm]{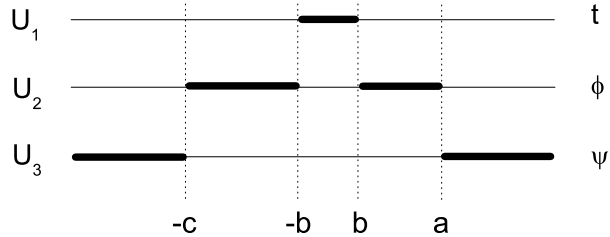}
\caption{Rod structure of static black ring on Kaluza-Klein bubbles} \label{rodstr_static}
        \end{center}
\end{figure}

The rod structure of the solution is depicted in Fig. \ref{rodstr_static}.  It is described as representing two Kaluza-Klein bubbles on a black ring.

As well known, the finite rod along the timelike Killing field corresponds to a horizon. In this case the horizon has $S^2\times S^1$ topology
where $S^1$ is not topologically supported -- consequently the black object is called a black ring. Similarly, the two finite spacelike rods denote the fixed point sets of the isometry generated by the Killing field associated with the compact dimension $\phi$. It can be proven that they represent two dimensional surfaces with disk topology. They are completely regular and, since they are spacelike, they restrict a region in spacetime which is causally disconnected from the rest of spacetime. For that reason such topological structures are called (spacetime) bubbles. Bubbles are spacetime structures with interesting properties. Frequently discussed feature is that they can separate several black objects, thus allowing the construction of multi-black hole spacetimes without conical singularities  \cite{IMT}, \cite{TIM2}, \cite{EHO}.

In the rest of the paper we will construct a rotating generalization of the described solution, endowing the black ring with a single angular momentum along the spacelike Killing field, which is associated with the non-compact dimension.

\section{Solution generation}

\subsection{Solution generation method}

Although it was possible to obtain the static solution describing two Kaluza-Klein bubbles on a black ring in a relatively simple way, the construction of its rotating counterpart will require more advanced solution generation methods. Fortunately, in 5-dimensional stationary axisymmetric spacetime all the solitonic techniques, developed for constructing solutions to the 4D stationary Einstein equations can be applied with certain modification. In the current paper, we preferred to use the auto-B\"{a}cklund transformation (BT) of the Ernst equation in the form invented by Neugebauer \cite{NG}, \cite{NG2} as most suitable for our purposes. We should note however that it is equivalent or closely related to a number of other transformations such as the inverse scattering transform, Hoenselaers-Kinnersley-Xantopoulos transform \cite{HKX} or Harrison transform \cite{Harrison}. Below, we will describe briefly the main idea of the solution generation method. We will use Weyl canonical coordinates $\rho$ and $z$, unless otherwise stated. The Killing fields are represented in adapted coordinates in the following way: $\xi=\partial/\partial {t}$ is the Killing field associated with time translations, $\eta=\partial/\partial \phi$ is the Killing field corresponding to the compact dimension and $\zeta=\partial/\partial \psi$ is the remaining spacelike Killing field.

\paragraph{} Without loss of generality, the 5D stationary axisymmteric metric with one axis of rotation can be represented in the Lewis-Papapetrou form
\begin{eqnarray}
ds^2 &=&-e^{2\chi-u}\left(dt-\omega d\psi\right)^2
       +e^{-2\chi-u}\rho^2d\psi^2
+e^{-2\chi-u}e^{2\Gamma}\left(d\rho^2+dz^2\right)
  +e^{2u}d\phi^2 \nonumber
\end{eqnarray}
where all the metric functions depend only on the Weyl canonical coordinates $\rho$ and $z$.

The 5D stationary axisymmetric gravitational field equations in vacuum can be presented in a convenient form using a complex potential $\cal E$. It corresponds directly to the potential introduced by Ernst in 4D \cite{Ernst}, its real and imaginary parts being connected to the metric functions $\chi$ and $\omega$ according to

\begin{eqnarray}\label{Ernst}
&&{\cal E} = e^{2\chi} + 2if ,
\end{eqnarray}
where $f$ is the twist potential defined by the equations

\begin{eqnarray}
&&\partial_{\rho}f= -{1\over 2}{e^{4\chi}\over \rho}\partial_{z}\omega ,\\
&&\partial_{z}f= {1\over 2}{e^{4\chi}\over \rho}\partial_{\rho}\omega .
\end{eqnarray}

The potential $\cal E$ provides an alternative description to the gravitational problem. Instead of using the metric functions (metric picture), the field equations can be written in terms of $\cal E$ and its complex conjugate $\cal E^*$  - the so called Ernst picture. Then the dimensionally reduced field equations yield the Ernst equation

\begin{eqnarray}
&&\left({\cal E} + {\cal E}^{*}\right)\left(\partial^2_{\rho} {\cal E} + \rho^{-1} \partial_{\rho}{\cal E}
+ \partial^2_{z} {\cal E}\right)= 2 \left(\partial_{\rho}{\cal E}\partial_{\rho}{\cal E}
+ \partial_{z}{\cal E}\partial_{z}{\cal E} \right). \nonumber
\end{eqnarray}

As it is easily noticed, in the static case when the twist potential vanishes, the nonlinear Ernst equation is simplified to the linear Laplace equation for the metric function $\chi$, which we discussed in section 2.

For any solution to the Ernst equation, the remaining metric function $\Gamma$ is determined again by a line integral

\begin{eqnarray}\label{Gammalin}
&&\rho^{-1}\partial_{\rho}\Gamma= {1\over \left({\cal E} + {\cal E}^{*} \right)^2}
\left[\partial_{\rho}{\cal E}\partial_{\rho}{\cal E}^{*} - \partial_{z}{\cal E}\partial_{z}{\cal E}^{*} \right] + \frac{3}{4}\left[\left(\partial_{\rho}u\right)^2 - \left(\partial_{z}u \right)^2\right],\nonumber \\
\nonumber \\
 &&\rho^{-1}\partial_{z}\Gamma=  \frac{2}{\left({\cal E} + {\cal E}^{*} \right)^2} \partial_{\rho}{\cal E} \partial_{z}{\cal E}^{*} + \frac{3}{2}~ \partial_{\rho}u \partial_{z}u.\nonumber
\end{eqnarray}

The solution of nonlinear partial  differential equations usually involves more refined analysis and complicated techniques. For certain classes of equations it is possible to find an auto-B\"{a}cklund transformation, which constitutes a set of relations that connect two different solutions to the equation. By the repetitive application of the B\"{a}cklund transformation different independent solutions can be obtained. Although these relations are much simpler than the original differential equation, they still involve integration. In some cases, however, there exist commutation theorems which can facilitate considerably the repetitive application of B\"{a}cklund transformations. They state that for any BT applied to a given solution there are three other transformations leading back to the same solution, and specify the relations between the parameters of the four BT that ensure the commutation. Once the commutation theorem has been proven, it allows the construction of solutions involving N subsequent applications of BT (N-soliton solutions) by purely algebraic transformations performed on a particular initial solution, called a seed.

As already mentioned Neugebauer has derived the auto-B\"{a}cklund transformation for the Ernst equation. He has also proven the corresponding commutation theorem, and deduced the explicit formulae for recursive calculation of BT's in its general form and in special cases. One of the most important cases for the applications is when B\"{a}cklund transformations are performed on a seed representing generalized Weyl solution. The rotating solution in the current article is also constructed in that manner. Therefore, in the following exposition we will outline the algebraic procedure of solution generation only for that particular case.

The B\"{a}cklund transformation for the Ernst equation is determined by a couple of functions $\lambda$ and $\alpha$, which satisfy the total Riccati equations

\begin{eqnarray}
d\lambda &=& \rho^{-1}(\lambda -1)\left[\lambda \rho_{, \zeta}d\zeta + \rho_{, \zeta^*}d\zeta^*\right], \nonumber \\
d{\cal \alpha} &=& \left({\cal E}_0 + {\cal E}^*_0 \right)^{-1}\left[({\cal \alpha} - \lambda^{1/ 2}){\cal E}^*_{0,\zeta}+ ({\cal \alpha}^2\lambda^{1/2}-{\cal \alpha}){\cal E}_{0,\zeta}\right]d\zeta + \nonumber \\
&&\left({\cal E}_0 + {\cal E}^*_0 \right)^{-1}\left[({\cal \alpha} - \lambda^{-1/2}){\cal E}^*_{0,\zeta^*}+ ({\cal \alpha}^2\lambda^{-1/2}-{\cal \alpha}){\cal E}_{0,\zeta^*}\right]d\zeta^*,
\end{eqnarray}
where ${\cal E}_0$ is the Ernst potential for the seed solution, $\zeta = \rho + iz$, the star denotes complex conjugation, and $(...),$ denotes differentiation.

If the seed solution belongs to the Weyl class, the coordinate dependence of the functions $\lambda$ and $\alpha$ is found to be

\begin{eqnarray}
\lambda_n &=& \frac{k_n - i\zeta^*}{k_n + i\zeta}, \nonumber \\
\alpha_{n}&=& {\mu_n + ie^{2\Phi_n}\over \mu_n - ie^{2\Phi_n}},
\end{eqnarray}
where $n$ denotes the number of the successive B\"{a}cklund transformation, $k_n$ and $\mu_n$ are real integration constants and $\Phi_n$ obeys the equation
\begin{eqnarray}\label{Riccatiphi}
d\Phi_n = \frac{1}{2} \lambda^{1/2}_n~ (\ln{{\cal E}_0})_{, \zeta}d\zeta + \frac{1}{2} \lambda^{-{1/2}}_n~ (\ln{{\cal E}_0})_{, \zeta^*}d\zeta^*.
\end{eqnarray}

The new solution to the Ernst equation obtained after $2N$ subsequent B\"{a}cklund transformations is constructed algebraically in the form

\begin{eqnarray}\label{ErnstWeyl}
{\cal E}= {\cal E}_0 {\det\left({{\cal \alpha}_{p}R_{k_p} - {\cal \alpha}_{q}R_{k_q} \over k_p - k_q } -1 \right)\over
\det\left({{\cal \alpha}_{p}R_{k_p} - {\cal \alpha}_{q}R_{k_q} \over k_p - k_q } +1 \right)}, ~~~\quad {\cal E}_0 = e^{2\chi_0},
\end{eqnarray}
where $p=1,3,...,2N-1$, $q=2,4,...,2N$, and the functions $R_{k_n}$ are given by
\begin{eqnarray}
R_{k_n}= \sqrt{\rho^2 + (z-k_n)^2}.
\end{eqnarray}

Further, we will consider only 2-soliton transformation. In this case the metric functions of the new solution can be expressed in explicit form taking into account ($\ref{Ernst}$)

\begin{eqnarray}
e^{2\chi}&=&e^{2\chi_0}\frac{W_1}{W_2} ,  \label{e^S} \\ \nonumber
\omega&=&e^{-2\chi_0}\frac{\hat{\omega}}{W_1}-C_\omega \label{omega} .    \\ \nonumber
\end{eqnarray}
The following functions are introduced \cite{Y3},

\begin{eqnarray}\label{metricfunc}
W_1&=&\left[(R_{k_1}+R_{k_2})^2-(\Delta k)^2\right] (1+ a b)^2  + \left[(R_{k_1}-R_{k_2})^2-(\Delta k)^2\right](a-b)^2  , \nonumber \\
W_2&=&\left[(R_{k_1}+R_{k_2}+\Delta k)+(R_{k_1}+R_{k_2}-\Delta k)a b \right]^2 \nonumber \\
 && +\left[(R_{k_1}-R_{k_2}-\Delta k)a - (R_{k_1}-R_{k_2}+\Delta k)b \right]^2 , \\ \nonumber  \\
{\hat \omega}&=& [(R_{k_1} + R_{k_2})^2-(\Delta k)^2](1+a b)\left[(R_{k_1}-R_{k_2} + \Delta k)b +
(R_{k_1}-R_{k_2} - \Delta k)a\right] \nonumber \\&& \,- [(R_{k_1} - R_{k_2})^2-(\Delta k)^2](b-a)
\left[(R_{k_1} + R_{k_2} + \Delta k) - (R_{k_1} + R_{k_2} - \Delta k)ab\right].  \nonumber
\end{eqnarray}
where we have denoted $\Delta k = k_2-k_1$.
The functions $a$ and $b$ are connected directly with $\Phi_n$, $n=1,2$ as

\begin{eqnarray}\label{abphi}
a = \mu_1^{-1}e^{2\Phi_1}, \quad  \bigskip b = -\mu_2 e^{-2\Phi_2}.
\end{eqnarray}
Taking into consideration these relations we will replace for convenience in the following discussion the real constants $\mu_1$ and $\mu_2$ with another pair $\alpha$ and $\beta$ defined as

\begin{eqnarray}
\alpha = \mu_1^{-1}, \quad  \bigskip \beta = -\mu_2.
\end{eqnarray}

As we have already obtained the explicit form of the Ernst potential for the new solution, we can solve the equations (\ref{Gammalin}) which determine the remaining metric function $\Gamma$. For the particular form of the Ernst potential (\ref{ErnstWeyl}) it can be expressed as

\begin{eqnarray}
e^{2\Gamma}= C {W_1 e^{2\gamma}\over \left(R_{k_{1}} +  R_{k_{2}}\right)^2 - (\Delta k)^2 },
\end{eqnarray}
Here $C$ is an integration constant and $\gamma$ is a solution to the linear system
\begin{eqnarray}\label{Gammalin2}
&&\rho^{-1}\partial_{\rho}\gamma = \left(\partial_{\rho}{\tilde \chi_{0}}\right)^2
- \left(\partial_{z}{\tilde \chi_{0}}\right)^2 + \frac{3}{4}\left[\left(\partial_{\rho}{u_0}\right)^2
- \left(\partial_{z}{u_0}\right)^2\right] , \nonumber \\ \\
&&\rho^{-1}\partial_{z}\gamma = 2 \partial_{\rho}{\tilde \chi_{0}} \partial_{z}{\tilde \chi_{0}} + \frac{3}{2} \partial_{\rho}{u_0} \partial_{z}{u_0},\nonumber
\end{eqnarray}
where we have used the auxiliary potential
\begin{eqnarray}
{\tilde \chi}_{0} = \chi_{0} +{1 \over 2} \ln{R_{k_1}+ R_{k_2} - \Delta k\over R_{k_1} + R_{k_2} +\Delta k}.
\end{eqnarray}

In conclusion, we will give the general formulae for applying a 2-soliton B\"{a}cklund transformation on a Weyl solution seed. From the presentation of the algebraic procedure it is obvious that in this case the B\"{a}cklund transformation is completely determined by the solution of the total Riccati equation (\ref{Riccatiphi}) and the linear system (\ref{Gammalin2}). Consider the general form of a Weyl seed

\begin{eqnarray}\label{seed}
{\cal E}_{0}= e^{2\chi_0}, \quad \bigskip \chi_{0}= \sum_{i} \varepsilon_i {\tilde U}_{\nu_i},
\end{eqnarray}
where $\varepsilon_i$ and $\nu_i$ are constants and the potentials ${\tilde U}_{\nu_i}$ have the form
\begin{eqnarray}
{\tilde U}_{\nu_i}={1\over 2} \left[R_{\nu_i}+ (z-\nu_i)\right] = {1\over 2} \ln\left[\sqrt{\rho^2 +(z-\nu_i)^2} + (z-\nu_i)\right].
\end{eqnarray}

Then, the total Riccati equation (\ref{Riccatiphi}) provided with the described type of Ernst potential has solution
\begin{eqnarray}\label{PhiWeyl}
\Phi_{n} = \sum_{i} {\varepsilon_i\over 2} \ln\left({e^{2U_{k_n}} + e^{2{\tilde U}_{\nu_i}}\over e^{{\tilde U}_{\nu_i}} }\right)
\end{eqnarray}
where  $U_{k_n}$, $n=1,2$ is defined by
\begin{eqnarray}
U_{k_n}= {1\over 2}\ln\left[R_{k_n} -(z-k_n)\right] .
\end{eqnarray}

The metric function $\gamma$ is obtained in a straightforward way from eqn. ($\ref{Gammalin2}$) as

\begin{eqnarray}\label{GammaWeyl}
\gamma = \gamma_{k_1,k_1} -2\gamma_{k_1,k_2} + \gamma_{k_2,k_2}  + 2\sum_{i}\left(\gamma_{k_1,\nu_i}-\gamma_{k_2,\nu_i} \right) +
\sum_{i,j}\varepsilon_i \varepsilon_j \gamma_{\nu_i,\nu_j}
\end{eqnarray}
where
\begin{eqnarray}
\gamma_{k,l}= {1\over 2}{\tilde U}_{k} + {1\over 2}{\tilde U}_{l} -{1\over 4} \ln\left[R_k R_l + (z-k)(z-l) +\rho^2  \right] .
\end{eqnarray}

We will use the working formulae we just presented for the generation of the solution describing rotating black ring on Kaluza-Klein bubbles.

\subsection{Seed solution}

\begin{figure}[h]
\begin{center}
      \includegraphics[width=8.cm]{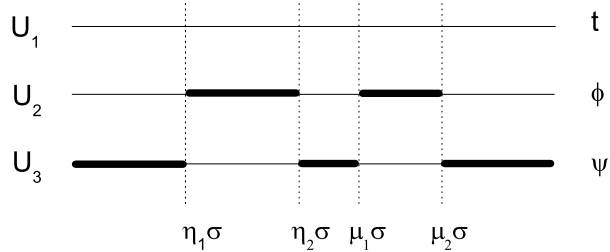}
\caption{Rod structure of the seed solution} \label{rodstr_seed}
        \end{center}
\end{figure}

It is observed \cite{IM} that B\"{a}cklund transformation with appropriately selected parameters is able to transform the solution in such a way that a finite rod along the axis of a spacelike Killing field is converted into rotational horizon with angular velocity along the same Killing vector, while the rest of the rod structure is preserved. For that reason a convenient seed for generation of rotating black ring on Kaluza-Klein bubbles is the generalized Weyl solution representing two static Kaluza-Klein bubbles in equilibrium.  The rod structure of the seed solution is presented in Fig. \ref{rodstr_seed}, where the finite rods $[\eta_1\sigma, \eta_2\sigma]$ and $[\mu_1\sigma, \mu_2\sigma]$ along the axes of the Killing field ${\partial/\partial \phi}$ correspond to the bubbles. The semi-infinite axes $( -\infty, \eta_1\sigma]$ and $[ \mu_2\sigma, \infty)$, and the finite rod $[ \eta_2\sigma, \mu_1\sigma]$ that separates the bubbles represent the fixed sets of the other spacelike Killing field ${\partial/\partial \psi}$. The rod endpoint parameters are aligned as $\eta_1\sigma < \eta_2\sigma < \mu_1\sigma < \mu_2\sigma$. As they are arbitrary, they can be chosen proportional to $\sigma$ for convenience in the subsequent solution generation procedure.

\noindent

We have obtained the metric functions of the seed solution in a straightforward way, following the method described in section 2,

\begin{eqnarray}\label{Seed}
&&g^0_{tt}= - 1, \nonumber \\
&&g^0_{\psi\psi}=\rho^2\frac{e^{2\widetilde{U}_{\eta_1\sigma}}e^{2\widetilde{U}_{\mu_1\sigma}}}{e^{2\widetilde{U}_{\eta_2\sigma}}
e^{2\widetilde{U}_{\mu_2\sigma}}},  \nonumber \\
&&g^0_{\phi\phi}= \frac{e^{2\widetilde{U}_{\eta_2\sigma}}e^{2\widetilde{U}_{\mu_2\sigma}}}{e^{2\widetilde{U}_{\eta_1\sigma}}
e^{2\widetilde{U}_{\mu_1\sigma}}}, \nonumber\\
&&g^0_{\rho\rho}= \frac{Y_{\eta_1\sigma, \eta_2\sigma}Y_{\eta_1\sigma, \mu_2\sigma}Y_{\eta_2\sigma, \mu_1\sigma}Y_{\mu_1\sigma, \mu_2\sigma}}{4 R_{\eta_1\sigma} R_{\eta_2\sigma} R_{\mu_1\sigma} R_{\mu_2\sigma}Y_{\eta_1\sigma, \mu_1\sigma}Y_{\eta_2\sigma, \mu_2\sigma}}
\frac{e^{2\widetilde{U}_{\eta_1\sigma}}e^{2\widetilde{U}_{\mu_1\sigma}}}{e^{2\widetilde{U}_{\eta_2\sigma}}
e^{2\widetilde{U}_{\mu_2\sigma}}},
\end{eqnarray}
where we have used the auxiliary functions
\begin{eqnarray}
R_{c}  &=& \sqrt{\rho^2 + (z-c)^2},  \\ \nonumber
\widetilde{U}_{c}&=&\frac{1}{2}\ln\left[R_{c}+(z-c)\right],  \quad  U_{c}=\frac{1}{2}\ln\left[R_{c}-(z-c)\right],\\ \nonumber
Y_{cd}&=& R_cR_d+(z-c)(z-d)+\rho^2.
\end{eqnarray}

In general, some regularity requirements arise on the rods along the spacelike Killing fields. In order to avoid the formation of conical singularities, the orbits of a spacelike Killing field should be identified with a certain period on all the rods in its direction, which is also compatible with the asymptotic structure of the spacetime. Taking into account that the length of the Kaluza-Klein circle at infinity is  $L_0$, the regularity conditions on the axes of the Killing field $\partial/\partial\phi$ corresponding to the compact dimension result in the relations

\begin{eqnarray}
(\Delta \phi)^0_{{\cal B}_i} = 2\pi \lim_{\rho\to 0}\sqrt{{\rho^2 g_{\rho\rho}\over g_{\phi\phi}}}= L_0,
\end{eqnarray}
where we have denoted the bubble rods with ${\cal B}_i$, $i = 1,2$.

Performing certain calculations, we can express the regularity conditions as
\begin{eqnarray}
(\Delta \phi)^0_{{\cal B}_1:~ \eta_1\sigma < z < \eta_2\sigma} &=& 4\pi\sigma(\mu_2 - \eta_1)\frac{\eta_2-\eta_1}{\mu_1 - \eta_1}, \\ \nonumber
(\Delta \phi)^0_{{\cal B}_2:~ \mu_1\sigma < z < \mu_2\sigma} &=& 4\pi\sigma(\mu_2- \eta_1)\frac{\mu_2-\mu_1}{\mu_2 - \eta_2}.
\end{eqnarray}
\noindent
The last expressions show that the solution is free of conical singularities with respect to $\phi$, if the two bubble rods possess equal length, i.e. if we restrict the rod endpoints values in such a way that $\eta_1\sigma = - \mu_2\sigma$ and $\eta_2\sigma = - \mu_1\sigma$.

In a similar way, we should identify the orbits of the other spacelike Killing field $\partial/\partial\psi$ on all the corresponding rods with its period at infinity $2\pi$,

\begin{eqnarray}
\Delta \psi = 2\pi \lim_{\rho\to 0}\sqrt{{\rho^2 g_{\rho\rho}\over g_{\psi\psi}}}=2\pi.
\end{eqnarray}
\noindent
On the semi-infinite axes the regularity condition is fulfilled identically. However, on the finite rod that separates both bubbles, it cannot be satisfied for any values of the rod structure parameters. Therefore, the generalized Weyl solution describing two static Kaluza-Klein bubbles possesses unavoidable conical singularity. By general considerations, it can be expected and interpreted intuitively as the force necessary to hold the bubbles apart in equilibrium. The presence of conical singularity in the seed solution will not affect the rotating solution we aim to generate, as we will see in the following sections.

\subsection{Rotating black ring on Kaluza-Klein bubbles}

In the generation of the solution describing rotating black ring on Kaluza-Klein bubbles we apply directly the algebraic procedure described in section 3.1., and more precisely, the working formulae deduced for the case of Weyl solution seed. We have performed the calculations in the general case with parameters of the two-soliton B\"{a}cklund transformation $k_1$ and $k_2$. However, the solution is invariant under translations in $z$-direction. For that reason we can set $k_1=\sigma$ and $k_2=-\sigma$  without loss of generality by performing the shift $z\rightarrow z + z_0$, where $z_0 = \frac{1}{2}(k_1 + k_2)$.  Below, we present the solution obtained by performing a two-soliton B\"{a}cklund transformation on the Weyl seed (\ref{Seed}). The functions $a$, $b$ and $\gamma$  are calculated according to (\ref{PhiWeyl}), (\ref{abphi}) and (\ref{GammaWeyl}) as

\begin{eqnarray} \label{ab}
a &=& \alpha
{e^{2U_{\sigma}} + e^{2{\widetilde U}_{\eta_2\sigma}} \over e^{{\widetilde U}_{\eta_2\sigma}} }
{e^{{\widetilde U}_{\eta_1\sigma}} \over e^{2U_{\sigma}} + e^{2{\widetilde U}_{\eta_1\sigma}}}
{e^{2U_{\sigma}} + e^{2{\widetilde U}_{\mu_2\sigma}} \over e^{{\widetilde U}_{\mu_2\sigma}}}
{e^{{\widetilde U}_{\mu_1\sigma}} \over e^{2U_{\sigma}} + e^{2{\widetilde U}_{\mu_1\sigma}}} , \\ \nonumber \\ \nonumber \\ \nonumber
b &=& \beta
{e^{2U_{-\sigma}} + e^{2{\widetilde U}_{\eta_1\sigma}} \over e^{{\widetilde U}_{\eta_1\sigma}} }
{e^{{\widetilde U}_{\eta_2\sigma}} \over e^{2U_{-\sigma}} + e^{2{\widetilde U}_{\eta_2\sigma}}}
{e^{2U_{-\sigma}} + e^{2{\widetilde U}_{\mu_1\sigma}} \over e^{{\widetilde U}_{\mu_1\sigma}}}
{e^{{\widetilde U}_{\mu_2\sigma}} \over e^{2U_{-\sigma}} + e^{2{\widetilde U}_{\mu_2\sigma}}} ,
\end{eqnarray}

\begin{eqnarray}
\gamma &=& \gamma_{\sigma, \sigma} - 2\gamma_{\sigma,-\sigma} + \gamma_{-\sigma,-\sigma}  - \left(\gamma_{\sigma,\eta_1\sigma} -\gamma_{-\sigma,\eta_1\sigma} \right) + \left(\gamma_{\sigma,\eta_2\sigma} - \gamma_{-\sigma,\eta_2\sigma} \right)  \nonumber \\ &&   -
\left(\gamma_{\sigma,\mu_1\sigma} -\gamma_{-\sigma,\mu_1\sigma} \right) + \left(\gamma_{\sigma,\mu_2\sigma} -\gamma_{-\sigma,\mu_2\sigma} \right)
 + \gamma_{\eta_1\sigma, \eta_1\sigma} + \gamma_{\eta_2\sigma, \eta_2\sigma} \nonumber \\ &&   +~ \gamma_{\mu_1\sigma, \mu_1\sigma} + \gamma_{\mu_2\sigma, \mu_2\sigma}  -2\gamma_{\eta_1\sigma, \eta_2\sigma} + 2\gamma_{\eta_1\sigma, \mu_1\sigma} - 2\gamma_{\eta_1\sigma, \mu_2\sigma} \nonumber \\ &&  -~2\gamma_{\eta_2\sigma, \mu_1\sigma} + 2\gamma_{\eta_2\sigma, \mu_2\sigma} - 2\gamma_{\mu_1\sigma, \mu_2\sigma} \\ \nonumber \\ \nonumber
\gamma_{cd}&=&\frac{1}{2}\widetilde{U}_{c}+\frac{1}{2}\widetilde{U}_{d}
           -\frac{1}{4}\ln [R_cR_d+(z-c)(z-d)+\rho^2]. \label{gamma}
\end{eqnarray}

\paragraph{}These functions completely determine the metric and we can write the new solution in the convenient form
\begin{eqnarray}\label{metric}
ds^2 &=& W~g^0_{tt}\left(dt - \omega d\psi \right)^2 + {g^0_{\psi\psi}\over W}d\psi^2 +
{Y\over W} g^0_{\rho\rho}\left(d\rho^2 + dz^2\right) +  g^0_{\phi\phi}d\phi^2, \\ \nonumber
{Y\over W} &=& Y_0\frac{W_2}{4R_\sigma R_{-\sigma}}\sqrt{\frac{Y_{\sigma, \eta_1\sigma}Y_{\sigma, \mu_1\sigma}Y_{-\sigma, \eta_2\sigma}Y_{-\sigma,  \mu_2\sigma}}{Y_{\sigma, \eta_2\sigma}Y_{\sigma,  \mu_2\sigma}Y_{-\sigma, \eta_1\sigma}Y_{-\sigma, \mu_1\sigma}}},  \\ \nonumber
W &=& \frac{W_1}{W_2}, \\ \nonumber
\end{eqnarray}
using the metric functions of the seed solution $g^0_{ij}$ derived in section 3.2. and the definitions of $W_1$ and $W_2$ (\ref{metricfunc}).

\noindent

\begin{figure} [h]
\begin{center}
      \includegraphics[width=8.cm]{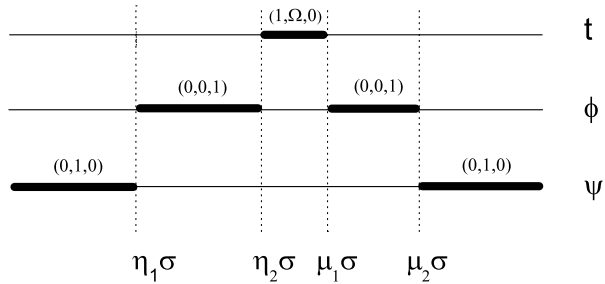}
\caption{Rod structure of rotating black ring on Kaluza-Klein bubbles} \label{rodstr}
        \end{center}
\end{figure}

The solution we obtained possesses the rod structure presented in Fig. \ref{rodstr}.  Again, the finite rods $[\eta_1\sigma, \eta_2\sigma]$ and $[\mu_1\sigma, \mu_2\sigma]$ along the axes of the Killing field ${\partial/\partial \phi}$ correspond to the bubbles, while the finite rod $[ \eta_2\sigma, \mu_1\sigma]$ separating the bubble is timelike and corresponds to a horizon. In contrast to the static solution \cite{Elvang}, however, it is directed along a linear combination of the Killing fields $v =  \partial/\partial t + \Omega~ \partial/ \partial \psi$ (denoted as $(1, \Omega, 0)$), which means that the horizon rotates with angular velocity $\Omega$ along the Killing field $\partial/\partial \psi$.
\bigskip

The described solution can be generated by two-soliton B\"{a}cklund transformation only if the transformation parameters  $k_1$ and $k_2$ are selected in a very specific way. Investigations show that the finite spacelike rod along the Killing vector $\partial/\partial \psi$ is transformed into a finite timelike rod directed as $(1, \Omega, 0)$ only if the rod structure and B\"{a}cklund transformation parameters are ordered as $-\sigma < \eta_2\sigma < \mu_1\sigma < \sigma$.  The  parameter ordering is further restricted as in general the two-soliton B\"{a}cklund transformation causes singular behavior of the metric function $W$ on some of the rods. It can be avoided (except at isolated points) again by appropriate parameter alignment, which in our case proved to be $\eta_1\sigma < -\sigma < \eta_2\sigma < \mu_1\sigma < \sigma <\mu_2\sigma$. Fortunately, it is compatible with the solution generation requirement just discussed.

Under the described ordering of the parameters, the metric function $W$ still remains singular at the bubble rod points $z = \sigma$ and $z=-\sigma$. The singularities can be removed, however, by imposing the following condition on the other two constants $\alpha$ and $\beta$, which were introduced by the B\"{a}cklund transformation,

\begin{eqnarray}\label{beta}
\alpha^2 &=& \frac{(1 - \eta_1)(1-\mu_1)}{(1-\eta_2)(\mu_2 - 1)}, \\ \nonumber
\beta^2 &=& - \frac{(1+ \eta_2)(1 +\mu_2)}{(1 + \eta_1)(1 + \mu_1)}.
\end{eqnarray}

Note that according to the parameter alignment $\eta_1 < -1 < \eta_2 < \mu_1 < 1 <\mu_2$, meaning that $\alpha$ and $\beta$, defined by the previous expression are real, as required.

\paragraph{} The solution contains another couple of integration constants $C_\omega$ and $Y_0$. They should be determined by the requirements to avoid global rotation of the spacetime and to preserve the Kaluza-Klein asymptotic, respectively.

\begin{eqnarray}\label{Y0}
C_\omega &=& \frac{4\sigma\beta}{1 + \alpha\beta},\quad \alpha + \beta = 0, \\ \nonumber
Y_0 &=& \frac{1}{(1 + \alpha\beta)^2}.
\end{eqnarray}
\noindent
Finally, we should ensure that the expressions, which determine $\alpha$ and $\beta$ are compatible with the condition $\alpha + \beta = 0$. It is easy to prove that this is possible only for rod structure parameters that fulfil the relations $\eta_1 = - \mu_2$ and $\eta_2=- \mu_1$.

\section{Analysis}

In this section we will examine the obtained solution. We will find its asymptotics, we will prove that it is free of conical singularities under certain requirements and we will compute its physical characteristics such as rotational velocity, temperature, mass and angular momentum. Finally, the Smarr relations are derived.

\subsection{Asymptotics}

We study the asymptotic behavior of the solution by introducing the asymptotic coordinates $r$ and $\theta$ defined as

\begin{eqnarray}
\rho&=&r\sin{\theta}, \nonumber \\
z&=&r\cos{\theta}. \nonumber
\end{eqnarray}

\noindent
In this way we receive the following asymptotic expansions for $r\longrightarrow\infty$

\begin{eqnarray}
g_{tt} &\approx& -1 + \frac{1 + \beta^2}{1-\beta^2} {2\sigma\over r} = -1 + \frac{c_t}{r},  \nonumber\\
g_{t\psi}&\approx& -\frac{2\sigma^2\beta \sin^2\theta}{(1-\beta^2)^2r}\left[\eta_1 -\eta_2 + \mu_1 - \mu_2 - 2 - \beta^2(\eta_1- \eta_2 + \mu_1 - \mu_2 + 2)\right] \nonumber \\
 &=& 2\frac{J}{L} \frac{\sin^2\theta}{r},  \nonumber\\
g_{\psi\psi} &\approx& r^2\sin^2\theta\left[1 - \frac{(\eta_1-\eta_2+\mu_1-\mu_2-2 - \beta^2(\eta_1-\eta_2+\mu_1-\mu_2+2))}{1 - \beta^2}{\sigma\over r}\right],  \nonumber\\
g_{\phi\phi} &\approx& 1 + \frac{(\eta_1 - \eta_2 + \mu_1 - \mu_2)\sigma}{r} = 1 +\frac{c_\phi}{r}, \nonumber \\
g_{\rho\rho} &\approx& 1 - \frac{(\eta_1 - \eta_2 + \mu_1 - \mu_2 - 2 - \beta^2(\eta_1 - \eta_2 + \mu_1 - \mu_2 + 2))\sigma}{(1-\beta^2)r}.
\end{eqnarray}

We see that up to leading order in the expansion, the solution asymptotically approaches the metric of the 5D Kaluza-Klein spacetime $R^4\times S^1$

\begin{equation}
ds^2 =  - dt^2 + dr^2 + r^2(d\theta^2 + \sin^2{\theta}d\psi^2) + d\phi^2.
\end{equation}

\subsection{Regularity conditions}

There exists a possibility the soliton transformation to introduce conical singularities on spacelike rods. Therefore, in order to get a regular solution we should ensure that the orbits of the Killing field  $\partial/\partial\phi$ on the bubble rods ${\cal B}_i$, $i = 1,2$ are identified with appropriate periods

\begin{eqnarray}
(\Delta \phi)_{{\cal B}_i} = 2\pi \lim_{\rho\to 0}\sqrt{{\rho^2 g_{\rho\rho}\over g_{\phi\phi}}}= L,
\end{eqnarray}
that coincide with the length of the Kaluza-Klein circle at infinity $L$.

In explicit form $\Delta \phi$ is expressed as follows
\begin{eqnarray}
(\Delta \phi)_{{\cal B}_i} = \sqrt{\frac{Y}{W}}|_{{\cal B}_i}(\Delta \phi)^0_{{\cal B}_i},
\end{eqnarray}
where $(\Delta \phi)^0_{{\cal B}_i}$ are the corresponding periods for the seed solution and the metric functions $Y/W$ are computed on the bubble rods
\begin{eqnarray}
\left(\frac{Y}{W}\right)_{{\cal B}_1:~ \eta_1\sigma < z < \eta_2\sigma} &=& -\frac{(1-\eta_1)}{(1+\eta_1)}\left[\frac{1 + \beta^2\frac{(1 + \eta_1)}{(1-\eta_1)}}{1 - \beta^2}\right]^2,  \nonumber \\
\left(\frac{Y}{W}\right)_{{\cal B}_2:~ \mu_1\sigma < z < \mu_2\sigma} &=& \frac{(\mu_2 + 1)}{(\mu_2 - 1)} \left[\frac{1 - \beta^2\frac{( \mu_2 - 1)}{(\mu_2 + 1)}}{1 - \beta^2}\right]^2.
\end{eqnarray}

The aforementioned conditions are compatible provided the bubble lengths coincide, i.e. we should set $\eta_1\sigma = - \mu_2\sigma$ and $\eta_2\sigma = - \mu_1\sigma$, as in the case of the seed solution. In the following analysis we adopt these relations between the rods structure parameters. Note that they agree exactly with the conditions that should be imposed on the expression for $\alpha$ and $\beta$ so that the requirement $\alpha + \beta = 0$ to be fulfilled. We can deduce the explicit form of the period $\Delta \phi$ taking into account (\ref{beta})

\begin{eqnarray}
\Delta \phi = 8\pi\frac{\mu_1\mu_2(\mu_2 - \mu_1)\sigma}{(\mu_1\mu_2 - 1)(\mu_2 + \mu_1)}\sqrt{(\mu_2^2 - 1)}.
\end{eqnarray}

In a similar way, we should identify the orbits of the other spacelike Killing field $\partial/\partial\psi$ on the semi-infinite axes with its period at infinity $2\pi$

\begin{eqnarray}
\Delta \psi = 2\pi \lim_{\rho\to 0}\sqrt{{\rho^2 g_{\rho\rho}\over g_{\psi\psi}}}=2\pi.
\end{eqnarray}
\noindent
It doesn't introduce further constraints, as it is fulfilled identically.

\subsection{Horizon}

As already mentioned the finite rod $[-\mu_1\sigma, \mu_1\sigma]$ corresponds to rotating horizon with $S^2\times S^1$ topology. Its angular velocity $\Omega$ can be computed by the requirement that the norm of the Killing vector $v =  \partial/\partial t + \Omega~ \partial/ \partial \psi$  vanishes on the horizon rod. Thus we obtain the following expression

\begin{equation}
\Omega = \frac{1}{2\sigma}\frac{\mu_1\mu_2 - 1}{\mu_1(\mu_2 - \mu_1)}\sqrt{\frac{(1-\mu^2_1)}{(\mu^2_2 - 1)}}.
\end{equation}

\begin{figure}[h]
\begin{center}
      \includegraphics[width=8.cm]{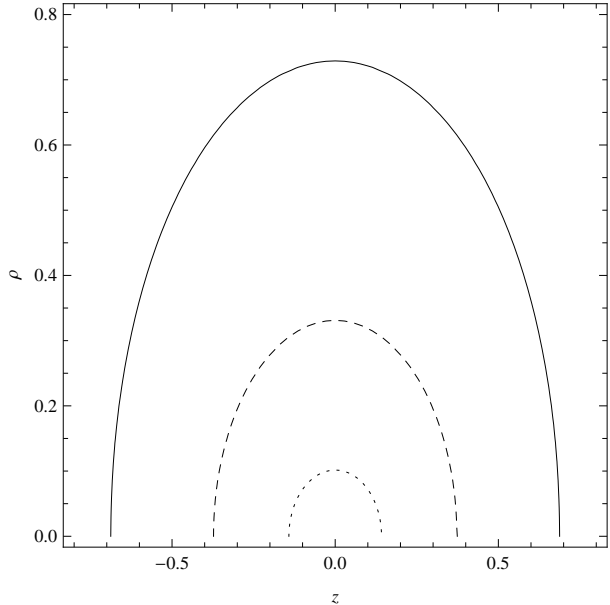}
\caption{Cross-section of the ergosphere in the $(\rho, z)$-plane for three different sets of values of the parameters $\mu_1$ and $\mu_2$, and $\sigma = 1$. The dotted line corresponds to $(\mu_1, \mu_2)=(0.1,1.1)$, the dashed line to $(\mu_1, \mu_2)=(0.2,1.2)$, and solid one to $(\mu_1, \mu_2)=(0.3,1.3)$. }\label{erg}
        \end{center}
\end{figure}

\paragraph{}In stationary spacetime an ergoregion exists defined as region in spacetime where the Killing vector $\partial/\partial t$ transforms from timelike to spacelike. It is bounded by a closed surface on which the metric function $g_{tt}$ vanishes. We will retain the term ergosphere introduced in 4D rotating solutions, although the surface does not have spherical topology. Investigations show that for our solution the ergosphere has the same $S^2\times S^1$ topology as the horizon and encompasses it completely without intersection points. The ergosphere intersects the $z$-axis at exactly two points, which are located symmetrically with respect to the horizon and lie on the bubble rods (Fig. \ref{erg}). They are determined by the expressions
\begin{eqnarray}
z =  \pm \mu_1\sigma\frac{1+\mu^2_1\mu^2_2 - 2\mu^2_2}{1+\mu^2_1\mu^2_2 - 2\mu_1^2}.
\end{eqnarray}
\noindent

Further, we have computed the area of the horizon
\begin{eqnarray}
A_{\cal H} = 32\pi L (\mu_1\sigma)^2 \frac{\mu_1\mu_2(\mu_2-\mu_1)(\mu_2^2-1)}{(\mu_1\mu_2- 1)^2(\mu_1+\mu_2)^2},
\end{eqnarray}
\noindent
It is directly connected with black ring's temperature as it can be observed by the relation derived in \cite{HY1}

\begin{eqnarray}
T_{\cal H} &=& L \frac{l_{\cal H}}{A_{\cal H}}. \\ \nonumber
\end{eqnarray}
Here $L$ is the length of the Kaluza-Klein circle at infinity and $l_{\cal H}$ is the length of the horizon rod.

Thus the temperature of the black ring is obtained
\begin{eqnarray}
T_{\cal H}&=& \frac{1}{16\pi\sigma}\frac{(\mu_1\mu_2-1)^2(\mu_1+\mu_2)^2}{\mu^2_1\mu_2(\mu_2-\mu_1)(\mu^2_2-1)} \\ \nonumber \\
&=& \frac{(\mu_1\mu_2-1)^2)(\mu_1+\mu_2)}{\mu_1(\mu_2-\mu_1)(\mu^2_2-1)}T^0_{\cal H}, \nonumber
\end{eqnarray}
where we have used the temperature of the static black ring on Kaluza-Klein bubbles $T^0_{\cal H}$ \cite{EHO}. It is expressed in our notations as

\begin{eqnarray}
T^0_{\cal H} &=& \frac{1}{16\pi}\frac{(\mu_1\sigma+\mu_2\sigma)}{\mu_1\sigma\mu_2\sigma}.
\end{eqnarray}

\subsection{Near horizon geometry}

Near the horizon of the black ring, i.e. in the limit $\rho\rightarrow0$, $-\mu_1\sigma < z < \mu_1\sigma$, the solution simplifies considerably. We introducing the Boyer-Lindquist coordinates R and $\theta$, familiar from the 4D Kerr solution and defined as (see e.g. \cite{Harmark}),

\begin{eqnarray}
\rho &=& \sqrt{R^2 - 2mR + \hat{a}^2}\sin{\theta}, \\ \nonumber
z &=& (R-m)\cos{\theta},
\end{eqnarray}
where $m$ is the mass of the Kerr black hole and $\hat{a}$ is the rotation parameter. Then the metric acquires the form

\begin{eqnarray}
ds^2 &=& - \frac{\Delta - {\cal A}^2(\theta)\sin^2{\theta}}{\Sigma'}\left( dt - \omega d\psi\right)^2 + \frac{f^2_{\mu_2\sigma}(\theta)}{f^2_{\mu_1\sigma}(\theta)}~\frac{\Sigma'}{\Delta - {\cal A}^2(\theta)\sin^2{\theta}}~\Delta \sin^2{\theta}d\psi^2+ \nonumber \\
&& + \frac{f^2_{\mu_1\sigma}(\theta)}{f^2_{\mu_2\sigma}(\theta)}d\phi^2+ C^2\Sigma'\left( \frac{dR^2}{\Delta} + d\theta^2 \right),
\end{eqnarray}
with the metric function $\omega$ is given by the expression

\begin{eqnarray}
\omega = \frac{f_{\mu_2\sigma}(\theta)}{f_{\mu_1\sigma}(\theta)}~2{\cal A}(\theta)\left[ 1 + M(\theta)\sin^2{\theta} \frac{R-m + M(\theta)}{\Delta - {\cal A}^2(\theta)\sin^2{\theta}}\right] - 2\hat{a}.
\end{eqnarray}

\paragraph{}We have defined the following functions depending only on $\theta$,

\begin{eqnarray}
f^2_{\mu_1\sigma}(\theta)&=& -a_{\sigma,\mu_1\sigma}.a_{\sigma,-\mu_1\sigma} + \sigma^2\sin^2{\theta}, \\ \nonumber
f^2_{\mu_2\sigma}(\theta)&=& a_{\sigma,\mu_2\sigma}.a_{\sigma,-\mu_2\sigma} + \sigma^2\sin^2{\theta},
\end{eqnarray}

\noindent as well as  the constants

\begin{eqnarray}
a_{\mu_i\sigma, \mu_j\sigma} = \mid\mu_i\sigma - \mu_j\sigma\mid, \quad ~~~
C &=& \frac{a_{\mu_2\sigma, -\mu_2\sigma}a_{\mu_1\sigma, -\mu_1\sigma}}{a_{\mu_1\sigma, -\mu_2\sigma}a_{\mu_2\sigma, -\mu_1\sigma}}\frac{a_{\sigma, -\mu_2\sigma}}{a_{\sigma, -\mu_1\sigma}}. \nonumber
\end{eqnarray}

\noindent The standard notations for the 4D Kerr solution in Boyer-Lindquist coordinates are used

\begin{eqnarray}
&&\Delta = R^2 - 2mR + \hat{a}^2, \quad ~~~ \Sigma = R^2 + \hat{a}^2\cos^2{\theta}, \\ \nonumber
&&\hat{a}= 2\sigma \frac{\beta}{1-\beta^2}, \quad ~~~ m = \sigma \frac{1+\beta^2}{1-\beta^2},
\end{eqnarray}
where $m$, $\hat{a}$ and $\beta$ are constants.

In a similar manner we have introduced the expressions

\begin{eqnarray}
\Sigma'= \left[R -m + M(\theta)\right]^2 + {\cal A}^2(\theta) \cos^2{\theta}, \\ \nonumber
{\cal A}(\theta) = 2\sigma \frac{b}{1-b^2}, \quad ~~~ M(\theta) = \sigma \frac{1+b^2}{1-b^2},
\end{eqnarray}
where $b$ is a function of $\theta$,

\begin{equation}
b = \beta \frac{a_{\sigma, -\mu_1\sigma}}{a_{\sigma, -\mu_2\sigma}}\frac{f_{\mu_2\sigma}(\theta)}{f_{\mu_1\sigma}(\theta)}.
\end{equation}

Note that the functions ${\cal A}(\theta)$ and $M(\theta)$ fulfil the same relation ${\cal A}^2(\theta) + M^2(\theta) = \sigma^2$, as the constants $\hat{a}$ and $m$.

In the limit $\mu_1\rightarrow 1$, $\mu_2\rightarrow 1$ the function $b(\theta)$ reduces to the constant $\beta$, and consequently $A(\theta)\rightarrow \hat{a}$, $M(\theta)\rightarrow m$ and $ \Sigma' \rightarrow \Sigma$.  It is easy to see that this limit procedure transforms the near-horizon metric into the rotating black string solution, as expected (see section 5.2.). The near horizon geometry can be interpreted as describing distorted rotating black string, which is deformed by the influence of the bubbles.

\subsection{Mass, tension and angular momentum}

The solution is characterized by three asymptotical charges - the ADM mass $M_{ADM}$, the tension ${\cal T}$ and the angular momentum ${\cal J}$, which are defined in the Komar integral formulation as

\begin{eqnarray}
M_{ADM} =  - {L\over 16\pi} \int_{S^{2}_{\infty}} \left[2i_\eta \star d\xi - i_\xi \star d\eta \right],
\end{eqnarray}

\begin{eqnarray}
{\cal T}= - {1\over 16\pi} \int_{S^{2}_{\infty}} \left[i_\eta \star d\xi - 2i_\xi \star d\eta \right],
\end{eqnarray}

\begin{eqnarray}
{\cal J} =  {L\over 16\pi} \int_{S^{2}_{\infty}} i_\eta \star d \zeta,
\end{eqnarray}
where $\star$  is the Hodge dual and $i_X$ is the interior product of the vector field $X$ with an arbitrary form.

Calculating the integrals we get the expressions

\begin{eqnarray}
M_{ADM} &=& \frac{L}{4}(2c_t - c_\phi) = \frac{L}{2}\sigma(\mu_2-\mu_1)\frac{(\mu_1\mu_2 + 1)}{(\mu_1\mu_2 - 1)}, \\ \nonumber
{\cal{T}}L &=& \frac{L}{4}(c_t - 2c_\phi) = \frac{L}{2}\sigma(\mu_2-\mu_1)\frac{(2\mu_1\mu_2 - 1)}{(\mu_1\mu_2 - 1)}, \\ \nonumber
{\cal J} &=& L \sigma^2~\frac{\mu_1\mu_2(\mu_2 - \mu_1)}{(\mu_1\mu_2 - 1)^2}\sqrt{(1-\mu^2_1)(\mu^2_2 - 1)}.
\end{eqnarray}

In addition to the asymptotic charges we can define local quantities - the intrinsic masses of the black ring and the bubbles

\begin{eqnarray}\label{BHKOMMARMASS}
M^{{\cal H}}= - {L\over 16\pi} \int_{{\cal H}} \left[2i_\eta \star d\xi - i_\xi \star d\eta \right],
\end{eqnarray}

\begin{eqnarray}\label{BKOMMARMASS}
M^{{\cal B}}_{i}= - {L\over 16\pi} \int_{{\cal B}_{i}} \left[2i_\eta \star d\xi - i_\xi \star d\eta \right],
\end{eqnarray}
where ${\cal H}$ is the 2-dimensional surface defined as the intersection of the horizon with a constant  $t$ and $\phi$
hypersurface and ${\cal B}_{i}$ $(i=1,2)$ are the bubble surfaces.

When calculations are performed these quantities get the simple form

\begin{eqnarray}
M_{\cal H}&=& L\sigma\frac{(\mu_2-\mu_1)}{(\mu_1\mu_2 - 1)}, \\ \nonumber
M_{{\cal B}_i}&=& \frac{L}{4}\sigma(\mu_2-\mu_1).
\end{eqnarray}
They satisfy as expected the following Smarr-like relations

\begin{eqnarray}
M_{ADM} &=& M_{\cal H} + \sum_i{M_{{\cal B}_i}}, \\ \nonumber
{\cal{T}}L &=&  \frac{1}{2}M_{\cal H} + 2\sum_i{M_{{\cal B}_i}}.
\end{eqnarray}

We can define in a similar manner a local angular momentum of the black ring of course

\begin{eqnarray}
{\cal J}_{\cal H} =  {L\over 16\pi} \int_{{\cal H}} i_\eta \star d \zeta.
\end{eqnarray}

It is of little significance by itself as in this case it coincides with the global charge $\cal J$. However, the Komar integral yields after some algebraic manipulations useful relation- the Smarr relation for the black ring

\begin{equation}
{\cal J} = \frac{1}{2\Omega}\left[ M_{\cal H} - \frac{L l_{\cal H}}{2}\right],
\end{equation}
which transforms into the more familiar form
\begin{equation}
M_{\cal H} = 2\left[{\cal J}\Omega + T S \right],
\end{equation}
when we take into consideration the definitions of the black ring's temperature $T$ and entropy $S$.

\section{Limit solutions}

\subsection{Static black ring on Kaluza-Klein bubbles}

The solution of Elvang and Horowitz is recovered in the particular case when the soliton transformation parameters $\sigma$ and $-\sigma$ coincide with the horizon rod endpoints. This is realized at the special value of the parameter $\mu_1 = 1$ at which $\beta$ vanishes as well, according to (\ref{beta}). Simple calculations show that in this limit the solution and its physical characteristics reduce to the static case \cite{Elvang}.

\subsection{Rotating black string}

In the limit when both Kaluza-Klein bubbles vanish the solution reduces to rotating black string. In order to take the limit in a proper way, we should shrink the bubbles to their inner points $z = \sigma$ and $z = -\sigma$, or, technically, let $\mu_1 \rightarrow 1$ and $\mu_2 \rightarrow 1$. Thus we obtain a single finite rod located at $-\sigma < z < \sigma$, which is timelike and corresponds to a black string. There is one more special feature we should consider taking the limit - the regularity conditions (\ref{beta}) should not be imposed on the parameters of the soliton transformation $\alpha$ and $\beta$. The metric function $W$ obtained after performing the described limit operation is not singular at $z = \sigma$ and $z = -\sigma$, but tends to zero instead. It corresponds exactly to the expected behavior since in this limit case the ergosphere intersects the $z$-axis at the horizon endpoints, so the metric function $g_{tt}= W$ should vanish at $z = \sigma$ and $z = -\sigma$.

\paragraph{} Performing the limit operation the seed solution (\ref{Seed}) reduces to the metric describing flat 4D spacetime trivially embedded in 5-dimensional spacetime. On the other hand, the functions (\ref{ab}) simplify to constants ($ a = \alpha$, $b = \beta$). In this way the solution acquires the familiar form of rotating black string (\cite{Mishima}, \cite{Kerr})

\begin{equation}
ds^2=-e^{2\Phi}\left( dt - \omega d\psi \right)^2+e^{-2\Phi}\left[
\rho^2 d\psi^2+e^{2\Lambda} \left( d\rho^2+ dz^2 \right)\right] + d\phi^2,
\end{equation}
where $\Phi$, $\Lambda$ and $\omega$ are the following functions

\begin{eqnarray}
\Phi &=&\frac{1}{2}\ln \left[ \frac{(R_{\sigma} + R_{-\sigma})^2-4m^2+ \hat{a}^2(R_{\sigma} - R_{-\sigma})^2/\sigma^2}
{(R_{\sigma} + R_{-\sigma} + 2m)^2+ \hat{a}^2(R_{\sigma} - R_{-\sigma})^2/\sigma^2}\right], \\ \nonumber
\Lambda &=& \frac{1}{2}\ln \left[ \frac{(R_{\sigma} + R_{-\sigma})^2-4m^2+ \hat{a}^2(R_{\sigma}-R_{-\sigma})^2/\sigma^2}{4R_{\sigma}R_{-\sigma}}\right], \\ \nonumber
\omega &=& \frac{\hat{a} m}{\sigma^2}\frac{(R_{\sigma} + R_{-\sigma}+2m)\left[ 4\sigma^2-(R_{\sigma}-R_{-\sigma})^2\right]}
{(R_{\sigma}+R_{-\sigma})^2-4m^2+\hat{a}^2(R_{\sigma}-R_{-\sigma})^2/\sigma^2}. \nonumber
\end{eqnarray}

Actually, this is a 4D Kerr solution trivially embedded in 5D spacetime, therefore we have presented the metric functions using the standard parameters for the 4D Kerr black hole - its mass $m$, and the rotation parameter $\hat{a}$ $(m^2 = \sigma^2 + \hat{a}^2)$. They can be expressed by means of $\beta$  and $\sigma$ as

\begin{eqnarray}
m &=& \frac{\sigma(1+\beta^2)}{1-\beta^2}, \quad \hat{a} =\frac{2\sigma\beta}{1-\beta^2}. \\ \nonumber
\end{eqnarray}

Note that we cannot deduce the physical characteristics of the limit solution directly from the corresponding expressions presented in the article, as we have performed all the calculations using the regularity conditions (\ref{beta}). However, if we start from the equivalent expressions

\begin{eqnarray}
\Omega &=& \frac{\beta(1-\beta^2)(1+\mu_1)}{2\sigma\left[(1-\beta^2)(1+\mu_2) + 2\beta^2(1+\mu_1)\right]}, \\ \nonumber
M &=& \frac{L\sigma\left[(1-\beta^2)(1+\mu_2) + 2\beta^2(1+\mu_1)\right]}{(1-\beta^2)\mu_1(1+\mu_2)}, \\ \nonumber
{\cal J} &=& \frac{2L\sigma^2\beta}{(1-\beta^2)^2}\left[(\mu_2 - \mu_1)(1-\beta^2)+(1+\beta^2)\right],
\end{eqnarray}
and apply the described limit procedure, we will get as expected (see Kerr solution, eg. \cite{Harmark})

\begin{eqnarray}
\Omega &=& \frac{\hat{a}}{2m(\sigma + m)}, \\ \nonumber
M &=& L \frac{\sigma(1+\beta^2)}{(1-\beta^2)}=Lm, \\ \nonumber
{\cal J} &=& L \frac{2\sigma^2\beta(1+\beta^2)}{(1-\beta^2)^2} = \hat{a} M,  \nonumber
\end{eqnarray}
\noindent
where $L$ is the length of the Kaluza-Klein circle at infinity.

\subsection{Boosted black string}

The solution describing boosted black string \cite{Emparan2} can also be obtained as a limit. In general it can be constructed by the 2-soliton transformation we described in section 3.1. from a seed

\begin{equation}\label{seed1}
ds^2 = - dt^2 + \frac{e^{2\widetilde{U}_{\mu_1\sigma}}}{e^{2\widetilde{U}_{\eta_2\sigma}}}d\psi^2 +\rho^2\frac{e^{2\widetilde{U}_{\eta_2\sigma}}}{e^{2\widetilde{U}_{\mu_1\sigma}}}d\phi^2 + \frac{Y_{\eta_2\sigma, \mu_1\sigma}}{2 R_{\eta_2\sigma} R_{\mu_1\sigma}}\frac{e^{2\widetilde{U}_{\eta_2\sigma}}}{e^{2\widetilde{U}_{\mu_1\sigma}}}\left(d\rho^2 + dz^2\right),
\end{equation}
where the functions $a$, $b$ and $Y$ are obtained by solving (\ref{Riccatiphi}) and (\ref{Gammalin2}) as

\begin{eqnarray} \label{abgamma}
a &=& \alpha
{e^{2U_{\sigma}} + e^{2{\widetilde U}_{\eta_2\sigma}} \over e^{{\widetilde U}_{\eta_2\sigma}}}
{e^{{\widetilde U}_{\mu_1\sigma}} \over e^{2U_{\sigma}} + e^{2{\widetilde U}_{\mu_1\sigma}}} {e^{U_{\sigma}}\over e^{{\widetilde U}_{\sigma}}} , \\ \nonumber
b &=& \beta
{e^{2U_{-\sigma}} + e^{2{\widetilde U}_{\mu_1\sigma}} \over e^{{\widetilde U}_{\mu_1\sigma}}}
{e^{{\widetilde U}_{\eta_2\sigma}} \over e^{2U_{-\sigma}} + e^{2{\widetilde U}_{\eta_2\sigma}}}{e^{{\widetilde U}_{-\sigma}}\over e^{U_{-\sigma}}}, \\ \nonumber
{Y\over W} &=& Y_0\frac{W_2}{4R_\sigma R_{-\sigma}}\sqrt{\frac{Y_{\sigma, \mu_1\sigma}Y_{-\sigma, \eta_2\sigma}}{Y_{\sigma, \eta_2\sigma}Y_{-\sigma, \mu_1\sigma}}}\frac{e^{2\widetilde{U}_\sigma}}{e^{2\widetilde{U}_{-\sigma}}}.  \\ \nonumber
\end{eqnarray}

The solution is regular if we impose the parameter ordering $ -\sigma < \eta_2\sigma < \mu_1\sigma < \sigma $, and the following restrictions on the constants $\alpha$ and $\beta$

\begin{eqnarray}\label{beta1}
\alpha^2 &=& \frac{(1-\mu_1)}{(1-\eta_2)}, \\ \nonumber
\beta^2 &=& \frac{(1+ \eta_2)}{(1 + \mu_1)}.
\end{eqnarray}

The constants $C_\omega$ and $Y_0$ are determined as

\begin{eqnarray}
C_\omega = 0, \quad ~~~ Y_0 = \frac{1}{(1 + \alpha\beta)^2}.
\end{eqnarray}

The boosted black string is recovered as a particular case of this class of solutions by setting $\eta_2\sigma = -\sigma$, or equivalently $\beta = 0$, and performing the coordinate transformation $z\rightarrow z + (\mu_1 + \eta_2)/2$.

From the solitonic generation of the boosted black string it is obvious that it can be obtained as a limit of the rotating black ring on Kaluza-Klein bubbles by setting $\mu_2 \rightarrow\infty$ and $\eta_1\rightarrow-\infty$ in appropriate way.\footnote{Note that no regularity conditions should be imposed on the solution parameters before performing the limit operations.} It means that we should ensure that both parameters tend to infinity in such a manner that their ratio remains finite, and further rescale the coordinates $\phi$ and $\psi$ such as

\begin{eqnarray}
\phi \rightarrow R \phi, \quad~~~ \psi \rightarrow \frac{\psi}{R},
\end{eqnarray}
by introducing a constant $R$, which tends to infinity at the same rate as $\mu_2\sigma$ and $\eta_1\sigma$ ($R/\mu_2\sigma$ and $R/\eta_1\sigma$ are finite). The performed rescaling corresponds actually to shifting the roles of the coordinates $\phi$ and $\psi$, so that $\psi$ describes the compact dimension now. It can be proven that in the described limit the seed solution (\ref{Seed}) reduces to the boosted black string seed (\ref{seed1}), and  the functions $a$, $b$ and $Y$ reduce to expressions (\ref{abgamma}).The regularity conditions (\ref{beta1}) should be imposed, as well as $\eta_2\sigma = -\sigma$, or $\beta = 0$, to recover the boosted black string.

\section{Discussion}

We presented an exact stationary axisymmetric solution to the 5D Einstein equations in spacetime with Kaluza-Klein asymptotic which describes rotating black ring on two Kaluza-Klein bubbles. The solution is extension to its static counterpart found by Elvang and Horowitz \cite{Elvang}. In contrast to the static solution, which can be constructed in a relatively simple way \cite{ER}, it is necessary to apply solitonic techniques to generate the rotating one. In the present article we have chosen to use two-soliton B\"{a}cklund transformation, although Belinski-Zaharov inverse scattering method is also relevant, if one is more familiar with it. A convenient seed solution to apply the B\"{a}cklund transformation is not the corresponding static case \cite{Elvang}, but a generalized Weyl solution representing two Kaluza-Klein bubbles held in equilibrium by a conical singularity. Despite, the generated solution is free of conical singularities. It possesses a horizon with $S^2\times S^1$ topology surrounded by an ergosphere that intersects the $z$-axis on the bubble rods. The rotation is performed in a single plane with angular velocity along the Killing vector $\partial/\partial\psi$. The mass, tension, angular velocity and angular momentum of the solution are computed and Smarr relations are derived. Finally, it is shown that some more simple solutions with the same symmetries can be obtained as particular limit cases of the generated solution, such as its static counterpart, the rotating black string and the boosted black string.

\section*{Acknowledgements}
We would like to thank Roberto Emparan for the useful suggestions and comments. The partial support by the Bulgarian National Science Fund under Grants  VUF-201/06, DO 02-257 and Sofia University Research Fund under Grant No 101/2010, is gratefully  acknowledged. P. Nedkova also acknowledges support from the European Operational programme HRD, contract BGO051PO001/07/3.3-02/53 with the Bulgarian Ministry of Education.

\end{document}